\documentclass[10pt,twocolumn,letterpaper]{article}

\usepackage[pagenumbers]{cvpr} 

\usepackage{graphicx}
\usepackage{amsmath}
\usepackage{amssymb}
\usepackage{booktabs}
\usepackage{bbm}
\usepackage{multirow}
\usepackage[ruled]{algorithm2e}
\usepackage[accsupp]{axessibility} 

\usepackage[pagebackref,breaklinks,colorlinks]{hyperref}

\usepackage[capitalize]{cleveref}
\crefname{section}{Sec.}{Secs.}
\Crefname{section}{Section}{Sections}
\Crefname{table}{Table}{Tables}
\crefname{table}{Tab.}{Tabs.}

\def\cvprPaperID{6790} 
\def\confName{CVPR}
\def\confYear{2022}

\usepackage{overpic}
\usepackage{enumitem} 
\usepackage{overpic} 
\usepackage{color}

\definecolor{turquoise}{cmyk}{0.65,0,0.1,0.3}
\definecolor{purple}{rgb}{0.65,0,0.65}
\definecolor{dark_green}{rgb}{0, 0.5, 0}
\definecolor{orange}{rgb}{0.8, 0.6, 0.2}
\definecolor{red}{rgb}{0.8, 0.2, 0.2}
\definecolor{darkred}{rgb}{0.6, 0.1, 0.05}
\definecolor{blueish}{rgb}{0.0, 0.3, .6}
\definecolor{light_gray}{rgb}{0.7, 0.7, .7}
\definecolor{pink}{rgb}{1, 0, 1}
\definecolor{greyblue}{rgb}{0.25, 0.25, 1}

\usepackage{blindtext}

\renewcommand{\paragraph}[1]{\vspace{1em}\noindent\textbf{#1}.}
\begin{document}
\title{CycleMix: A Holistic Strategy for Medical Image Segmentation from\\Scribble Supervision}

\author{Ke~Zhang and~Xiahai Zhuang\thanks{Xiahai Zhuang is corresponding author. This work was funded by the National Natural Science Foundation of China (grant no. 61971142, 62111530195 and 62011540404) and the development fund for Shanghai talents (no. 2020015)}\\
Fudan University\\}
\maketitle
\begin{abstract}
Curating a large set of fully annotated training data can be costly, especially for the tasks of medical image segmentation.
Scribble, a weaker form of annotation, is more obtainable in practice, 
but training segmentation models from limited supervision of scribbles is still challenging.
To address the difficulties, we propose a new framework for scribble learning-based medical image segmentation, which is composed of mix augmentation and cycle consistency and thus is referred to as \emph{CycleMix}.
For augmentation of supervision, CycleMix adopts the mixup strategy with a dedicated design of random occlusion, to perform increments and decrements of scribbles.
For regularization of supervision, CycleMix intensifies the training objective with consistency losses to penalize inconsistent segmentation, which results in significant improvement of segmentation performance.  
Results on two open datasets, \textit{i.e.}, ACDC and MSCMRseg, showed that the proposed method achieved exhilarating performance, demonstrating comparable or even better accuracy than the fully-supervised methods. 
The code and expert-made scribble annotations for MSCMRseg are publicly available at \url{https://github.com/BWGZK/CycleMix}.
\end{abstract}
\section{Introduction}
\label{sec:intro}
Large fully-annotated datasets are crucial to the generalization ability of deep neural networks. 
However, the manual labeling of medical images requires great efforts from experienced clinical experts, which is both expensive and time-consuming.
To alleviate it, existing works have exploited weakly labeled and unlabeled training data to assist model training, such as semi-supervised learning (SSL) \cite{NIPS2017_68053af2,souly2017semi,mittal2019semi} and weakly-supervised learning (WSL) \cite{wei2016stc,pathak2015constrained,khoreva2017simple}.
However, SSL generally requires part of the images in the dataset to be accurately and precisely annotated.
As an alternative, we propose to investigate a specific form of WSL approaches, which only utilize scribble annotations for model training.

\begin{figure}
     \centering
     \begin{subfigure}[b]{\linewidth}
         \centering
         \includegraphics[width=0.96\linewidth]{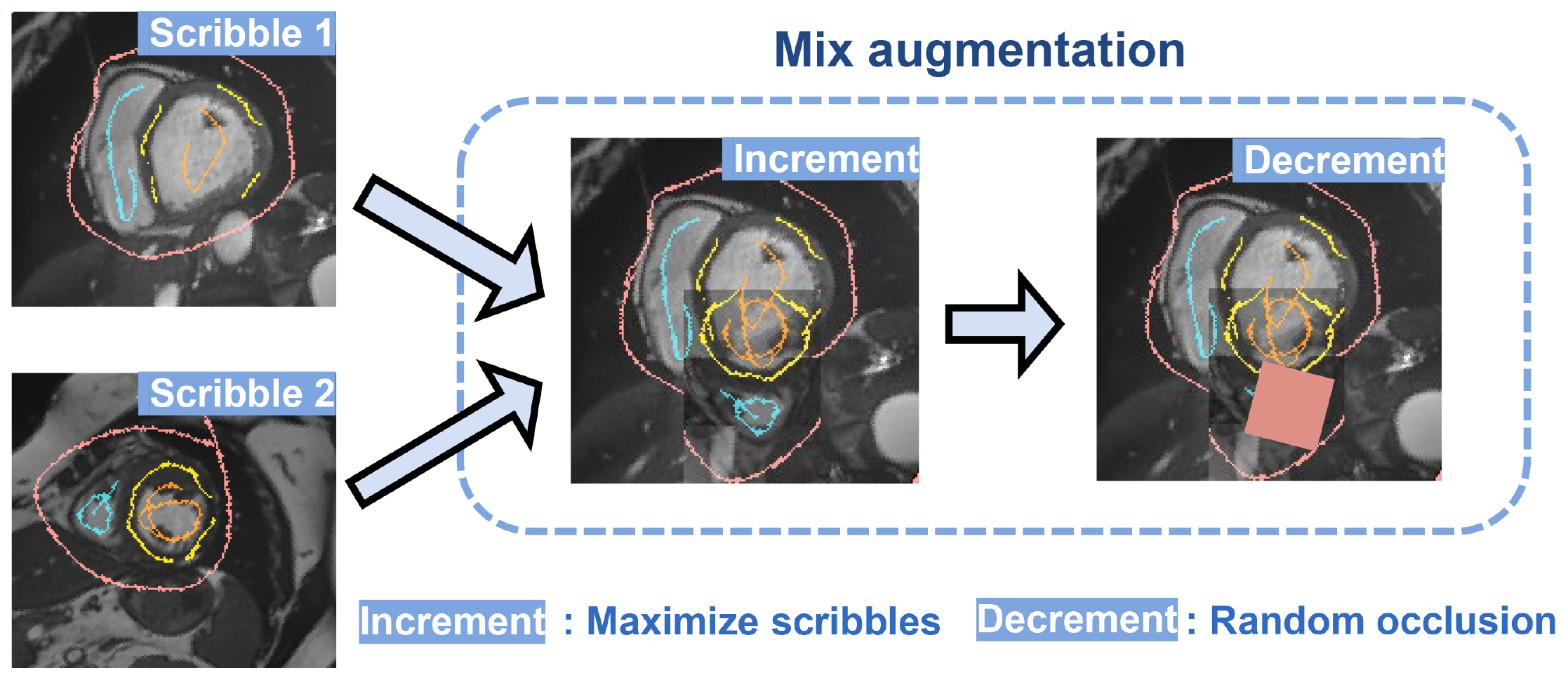}
         \caption{}
     \end{subfigure}
     \begin{subfigure}[b]{\linewidth}
         \centering
         \includegraphics[width=\linewidth]{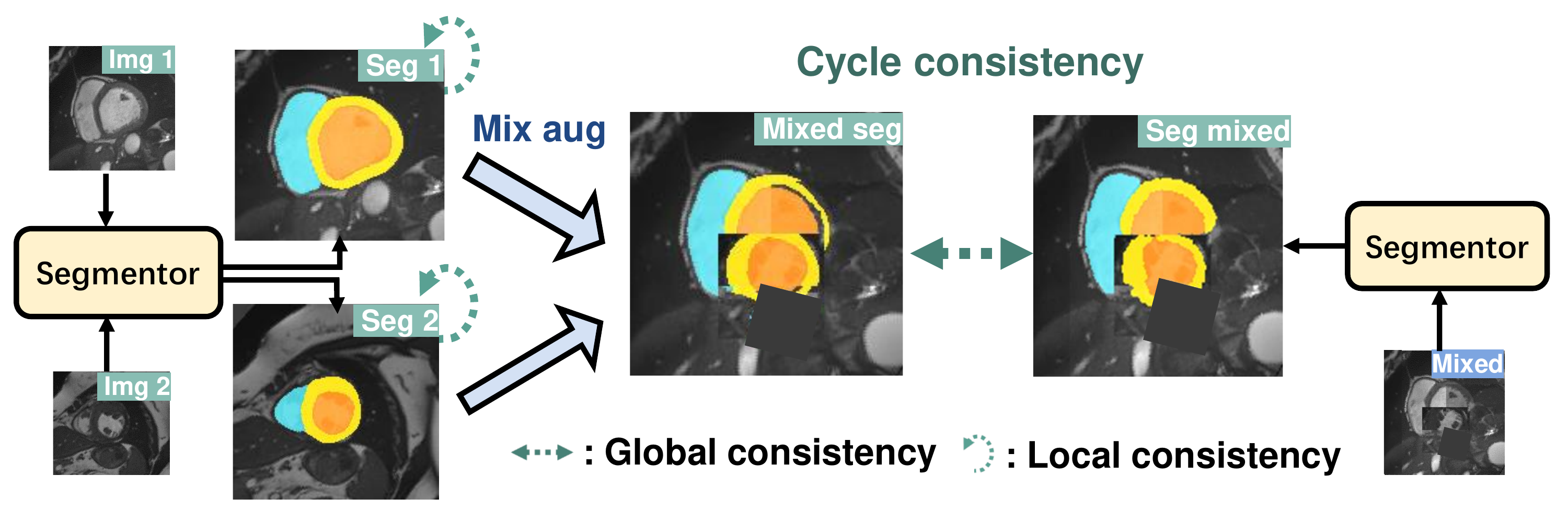}
         \caption{}
         \label{fig:three sin x}
     \end{subfigure}
\caption{Illustration of CycleMix based on mix augmentation and cycle consistency on scribble training images: 
(a) shows the operations and results of mixing images and corresponding labels;  
(b) illustrates the segmentation results for consistency regularization. 
}
\label{fig:illustration}
\end{figure}

WSL is proposed to exploit weak annotations, such as image-level labels, sparse annotations, and noisy annotations~\cite{tajbakhsh2020embracing}.
Among them, scribble, as images in Figure.~\ref{fig:illustration} (a) illustrate, is one of the most convenient forms of weak label and has great potential in medical image segmentation~\cite{Can2018LearningTS}.
However, due to the lack of supervision, it is still ardours to learn the shape priors of objects, which makes the segmentation of the boundaries particularly difficult.

The existing scribble learning mainly includes two groups. 
The first line of researches leverage \textit{a priori} assumption to expand scribble annotation~\cite{tajbakhsh2020embracing}, such as labeling pixels with similar gray values and similar positions in the same category \cite{l2016inscribblesup,ji2019scribble}.
However, the process of scribble expansion may generate noisy labels, which deteriorates the segmentation performance of trained models.
The second one learns adversarial shape priors, but requires extra fully-annotated masks~\cite{9389796,Larrazabal2020PostDAEAP,zhang2020accl}.

There is a line of augmentation strategies, well known as \emph{mixup}, have been proposed, which focus on generating previously-unseen virtual examples~\cite{zhang2018mixup,devries2017cutout,yun2019cutmix,kimICML20,kim2021comixup}.
However, these strategies are proposed for image classification,
and they may change the shape priors of target objects, leading to unrealistic segmentation results for a segmentation task.
When only scribble supervision is available, the segmentation performance using mixup augmentation could become even worse and unstable, due to the lack of precise annotations.



To address above mentioned challenges, we propose \textit{CycleMix} to learn segmentation from scribbles.
As illustrated in Figure.~\ref{fig:illustration}, CycleMix maximizes supervision of scribbles based on mix augmentation and random occlusion, and regularizes training of models using  consistency losses. 
Firstly, we surmise that a segmentation model should benefit from finer gradient flow via larger portion of annotated pixels.
Therefore, we propose the two-step \textit{mix augmentation} strategy to augment supervision, including image combination to increase scribbles and random occlusion to reduce scribbles.
In addition, we develop two-level \textit{consistency} regularization, at both of the global and local levels.
The global consistency loss penalizes the inconsistent segmentation of the same image patch in two scenarios, \textit{i.e.}, in the original image and mixed image; while
the local consistency loss minimizes the distance between prediction and its largest connected component, 
exploiting the prior knowledge of anatomy that the target structures are interconnected.

The contributions of this paper are summarized as follows:
\begin{itemize}
\setlength\itemsep{-.3em}
\item We propose a novel weakly-supervised segmentation framework for scribble supervision, \textit{i.e.}, CycleMix, by integrating mix augmentation of supervision and regularization of supervision from consistency, and introduce a new scribble annotated cardiac segmentation dataset of MSCMRseg.
\item To the best of our knowledge, the proposed CycleMix is the first framework to incorporate mixup strategies for augmentation of weakly-supervised segmentation, where one can achieve both increments and decrements of scribbles from the mixed training images.
\item We propose the consistency losses to regularize the limited supervision from scribbles by penalizing inconsistent segmentation results, at both the global and local levels, which can lead to profound improvement of model performance.
\item CycleMix has been evaluated on two open datasets, \textit{i.e.}, ACDC and MSCMR, and demonstrated promising performance by generating comparable or even  better segmentation accuracy than the fully-supervised approaches.

\end{itemize}

\section{Related works}

\subsection{Learning from scribble supervision}
Scribble refers to sparse annotations where masks are provided for a small fraction of pixels in images~\cite{tajbakhsh2020embracing}.
Existing methods mostly used selective pixel loss for annotated pixels.
There are works \cite{bai2018recurrent,l2016inscribblesup,ji2019scribble} attempting to expand scribbles or reconstruct the complete mask for model training.
However, the pixel-relabeling process required iterative training, which is slow and prone to noisy labels.
To avoid relabeling, several works utilized conditional random field  to refine the segmentation results in post-processing \cite{chen2017deeplab,Can2018LearningTS} or as trainable layer \cite{zheng2015conditional,Tang2018OnRL}.
However, these methods could not provide better supervision for model training.
Other works \cite{9389796,zhang2020accl} included a new module to evaluate the quality of segmentation masks, which encourages the predictions to be realistic.
For example, Gabriele \textit{et al.} \cite{9389796} proposed the multi-scale attention gates in adversarial training, Zhang \textit{et al.} \cite{zhang2020accl} used PatchGAN discriminator~\cite{isola2017image} to leverage shape priors.
However, these methods required additional data source of complete masks.

\subsection{Mixup augmentations}
Data augmentation plays a vital role in preventing models from overfitting to the limited training data and enhancing the generalization ability of neural networks.
Mixup augmentations refer to a line of strategies which combine two images and corresponding labels~\cite{zhang2018mixup,devries2017cutout,yun2019cutmix,kimICML20,kim2021comixup}. 
Compared with conventional augmentation methods, \textit{i.e.}, rotation and flipping, mixup approaches can increase scribble annotations of augmented image through mix operation.
Zhang \textit{et al.} \cite{zhang2018mixup} introduced MixUp, which performed linear interpolation between two images and their labels. 
Manifold MixUp in \cite{verma2019manifold} extended the mixup operation of input images to hidden features.
Cutout in \cite{devries2017cutout} randomly dropped out the square regions of images, and CutMix in \cite{yun2019cutmix} replaced the dropped areas with patches from other images.
Puzzle Mix in \cite{kimICML20} introduced a new mixup method based on saliency and local statistics.
Co-mixup in \cite{kimICML20} extended the mixup between two images to multiple images, and encouraged the supermodular diversity of mixed images.

In medical imaging, mixup augmentation has been applied to semi-supervised image segmentation \cite{chaitanya2019semi} and object detection tasks \cite{wang2020focalmix}. Chaitanya \textit{et al.} \cite{chaitanya2019semi} concluded that mixup could lead to an impressive performance gain on semi-supervised segmentation.
Although the mixed images might not look realistic, the mixed soft labels can provide more information to facilitate the training of models \cite{chaitanya2019semi,hinton2015distilling}.

\begin{figure*}[t]
\centering
\includegraphics[width=0.96\linewidth]{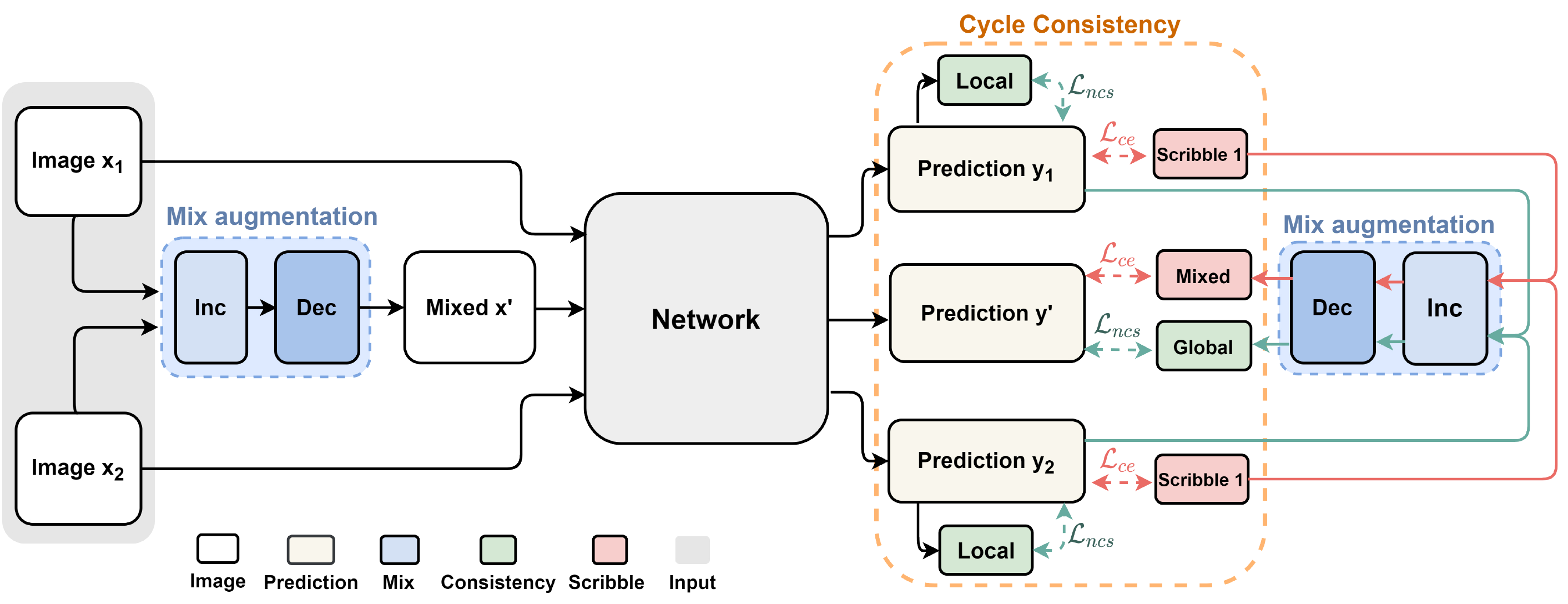}
\caption{The pipeline of CycleMix, where two blue dashed boxes represent the same mix augmentation operation.
The loss functions include the losses of scribble supervision (annotated with red) and the losses of consistency (annotated with green).
}
\label{fig:teaser}
\end{figure*}

\subsection{Consistency regularization}
Consistency strategies take advantage of the fact that if the same image is perturbed, the segmentation results should remain consistent.
Consistency regularization has been widely applied in image-translation and semi-supervised learning.
CycleGAN~\cite{zhu2017unpaired} leveraged forward-backward consistency to enhance the ability of image-to-image translation.
In semi-supervised setting, consistency is enforced over two augmentation versions of input images to obtain stable predictions of unlabeled images~\cite{laine2016temporal,NIPS2017_68053af2,ouali2020semi}.
In this work, we propose to utilize consistency at both of the global and local levels to leverage the mix-invariant property and interconnected fact of segmentation structures.


\section{Method}
The proposed CycleMix is composed of two new strategies, \textit{i.e.}, mix augmentation of scribble supervision and cycle consistency for regularization of supervision. 
The former is aimed to achieve the increments and decrements of scribbles by two-step mix-based image combination and random occlusion;
the latter is designed to regularize the supervision in model training via two-level consistency penalty.
Figure.~\ref{fig:teaser} presents framework of neural network implementation of CycleMix.

\subsection{Mix augmentation of scribble supervision}
In this section, we extend the mixup strategy to the two-stage augmentation of scribble supervision. 
In the first stage, we increase the amount of scribbles by image combination, referred to increments of scribbles, which is to mixup two images to maximize the saliency.
In the second stage, we perform an operation of random occlusion, by replacing certain area containing scribbles with background, which results in decrements of scribbles.
Finally, the augmentation of supervision is achieved via a dedicated loss function from the generated mixup images.

\subsubsection{Increments of scribbles}
We surmise that increasing scribbles will benefit from finer gradient flow through larger proportions of annotated pixels.
Furthermore, we observe that the scribble-annotated area generally has high saliency.
Therefore, we propose to maximize the scribble annotation of mixed images to efficiently obtain the maximization of saliency of mixed training images.
Here, we adopt the Puzzle Mix in \cite{kimICML20} to utilize saliency and local statistic features. 
Note that the proposed method is applicable to other mixup strategies, such as MixUp~\cite{zhang2018mixup}, CutMix~\cite{yun2019cutmix} and Co-mixup~\cite{kim2021comixup}. 
Readers could refer to the supplementary material for a comparison study.

We apply Puzzle Mix to both images and their corresponding scribble labels.  
Let two $d$-dimensional images with annotations be $(x_1, y_1)$, $(x_2, y_2)$. The mixed result transported from the two training data, denoted as $(x^m_{12},y^m_{12})$, is computed by:
\begin{gather}
    x^m_{12} = M(x_1, x_2)\ \text{\ and\ }\ y^m_{12} = M(y_1, y_2), \label{eq:miximage}\\
    M(a_1, a_2) = (1-z)\odot \textstyle\prod_1^T a_1+ z\odot \textstyle\prod_2^T a_2, \label{eq:mixop}
\end{gather}
where $M(a_1, a_2)$ is the mixup function on $a_1$ and $a_2$; $\scriptstyle\prod_1$ and $\scriptstyle\prod_2$ represent the transportation matrix of dimension $d\times d$; $z$ denotes a mask in $[0, 1]$ of dimension $d$; $\odot$ refers to the element-wise multiplication. 
The parameter set, $\{{\scriptstyle\prod_1}, {\scriptstyle\prod_2},z\}$, is aimed to maximize the saliency of mixed image, which is computed by,
$$
\{{\scriptstyle\prod_1}, {\scriptstyle\prod_2},z\} = 
\underset{{\scriptstyle\prod_1},{\scriptstyle\prod_2}, z}{\arg\max}[ (1-z)\odot {\scriptstyle\prod_1^T} s(x_1) + z\odot {\scriptstyle\prod_2^T} s(x_2) ],
$$
where $s(x)$ is the saliency of image $x$ and is computed by taking the $l_2$ norm of the gradient value. 
For this optimization, one could refer to \cite{kimICML20} for more details.

\subsubsection{Decrements of scribbles}
To further augment scribble supervision, we propose to randomly occlude a region containing scribbles from the mixed images, to generate more training images.  
This strategy results in decrements of scribbles in the mixed image, and has been proved to be effective in enhancing performance of object localization~\cite{yun2019cutmix}.

Let $(x^o, y^o)$ be the pair of new training data generated from $(x^m,y^m)$.
We apply a randomly rotated rectangular area to occlude the image and turns the occluded scribbles into background,  
\begin{gather}
    x^o_{12} = (1-\mathbbm{1}_O) \odot x^m_{12}\\
    y^o_{12} = (1-\mathbbm{1}_O) \odot y^m_{12}
\end{gather}
where $\mathbbm{1}_O$ is a binary rectangular mask of dimension $n\times n$.
In our experiment, we chose a rectangle with size of $32\times32$.

\subsubsection{Scribble supervision}
For scribble supervision, we apply the cross-entropy function \emph{solely on the annotated pixels, ignoring the unlabeled pixels whose ground truth labels are unknown}.
Hence, the loss $\mathcal{L}_{unmix}$ for unmixed samples $(x_1,y_1)$ and $(x_2, y_2)$ is formulated as:
\begin{equation}
    \mathcal{L}_{unmix} = \frac{1}{2}\left[\mathcal{L}_{ce}(\hat{y}_1,y_1)+ \mathcal{L}_{ce}(\hat{y}_2, y_2)\right],
\end{equation}
where, $\hat{y}=S(x)$ is the predicted segmentation of $x$, and,
\begin{equation}
\mathcal{L}_{ce}(\hat{y},y) = \sum_{i\in\Omega_L} \sum_{k\in K}-y[i,k]\log(\hat{y}[i,k]),
\end{equation} 
where, $K$ is the index set of labels, $[i,k]$ indicate the $k$-element of label vector of the $i$-th pixel, 
$y[i,k]$ equals the probability of $i$-th pixel belongs to the $k$-th class,
and $\Omega_L$ refers to the set of pixels with scribble annotation, to which $\mathcal{L}_{ce}$ loss is applied.

Furthermore, since the operation of Puzzle Mix is not symmetric, namely $M(x_1,x_2)\neq M(x_2,x_1)$, we use a symmetrical loss, referred to as mixed loss $\mathcal{L}_{mix}$, for the generated samples $(x^o_{12}, y^o_{12})$ and $(x^o_{21}, y^o_{21})$, 
\begin{equation}
\begin{aligned}
    \mathcal{L}_{mix} &= \frac{1}{2}\left[\mathcal{L}_{ce}(\hat y ^o_{12},y^o_{12})+\mathcal{L}_{ce}(\hat y ^o_{21},y^o_{21})\right].\\
\end{aligned}
\end{equation}

The loss for augmented scribble supervision is given by, 
\begin{equation}
\mathcal{L}_{sup} = \lambda_1\mathcal{L}_{unmix} + \lambda_2\mathcal{L}_{mix}, 
\end{equation}
where $\lambda_1$,$\lambda_2$ are the balancing parameters.

\subsection{Regularization of supervision via cycle consistency}
In this section, we introduce two regularization terms, \textit{i.e.}, the global consistency loss and the local consistency loss.

\subsubsection{Global consistency}
The objective of global consistency is to leverage the mix-invariant property, which requires the same image patch to behave consistently in two scenarios, \textit{i.e.}, the original image and the mixed image.
Therefore, we propose the global consistency loss to penalize the inconsistent segmentation.

For images $x_1,x_2$ and their mixed image $x^m_{12}=M(x_1, x_2)$, the corresponding segmentation is represented as $\hat{y}=S(x)$, where $S(\cdot)$ is the segmentor.
Assume the parameters of mixing function, \textit{i.e.},  $\scriptstyle\prod_1$, $\scriptstyle\prod_2$, and $z$ in in Eq.~(\ref{eq:mixop}), remain unchanged, one should have,
\begin{equation}
    M(S(x_1), S(x_2)) = S(M(x_1, x_2)).
\label{3.1eq1}
\end{equation}
This is the global consistency requiring the mixed segmentation of image $x_1$ and $x_2$ to be consistent with the segmentation of the mixed image $x^m_{12}$ after the same mixing operation.
Taking the random occlusion operation into account, we modify Eq.~(\ref{3.1eq1}) as follows,
\begin{equation}
    (1-\mathbbm{1}_O)\odot M\left(\hat{y}_1, \hat{y}_2\right) 
    = 
    S\left((1-\mathbbm{1}_O)\odot x^m_{12}\right).
\label{4.1eq2}
\end{equation}
We propose to use a symmetrical metric based on the negative cosine similarity between two segmentation results as the global consistency loss \cite{chen2020simsiam,grill2020bootstrap},
\begin{equation}
    \mathcal{L}_{con\text{-}g} = \frac{1}{2}\left[\mathcal{L}_{ncs}(p_{12},q_{12}) + \mathcal{L}_{ncs}(p_{21}, q_{21})\right],
\label{4.1eq4}
\end{equation}
where, 
$p_{12} \triangleq (1-\mathbbm{1}_o)\odot M(\hat{y}_1, \hat{y}_2)$ and $q_{12} \triangleq S((1-\mathbbm{1}_o)\odot x^m_{12})$ are respectively the mixed segmentation and segmentation of mixed image, and likewise for $p_{21}$ and $q_{21}$; 
$\mathcal{L}_{ncs}(\cdot,\cdot)$ is the negative cosine similarity and is defined as,
\begin{equation}
    \mathcal{L}_{ncs}(p, q) = -\frac{p\cdot q}{||p||_2 \cdot ||q||_2}.
\end{equation}

\subsubsection{Local consistency}
For a target object, the mixup operation often causes disconnected structure in the mixed image.
This phenomenon makes it particularly difficult for a segmentation model to learn the shape priors of target objects.

Leveraging the fact that the target structure can be interconnected in many medical applications, we propose the local consistency to eliminate the discrete results. For unmixed images $x_1$ and $x_2$, the local consistency loss $\mathcal{L}_{con\text{-}l}$ is formulated as: 
\begin{equation}
\small
    \mathcal{L}_{con\text{-}l} = \frac{1}{2}\left[\mathcal{L}_{ncs}\left(\hat{y}_1, C\left(\hat{y}_1\right)\right) + \mathcal{L}_{ncs}\left(\hat{y}_2, C\left(\hat{y}_2\right)\right)\right],
\label{4.2eq1}
\end{equation}
where, $C(\cdot)$ is a morphological operation on a segmentation result, which outputs the largest connected area of each non-background class in the input segmentation.
The purpose of Eq.~(\ref{4.2eq1}) is to minimize the distance between segmentation results and their largest connected areas. As formulated in Eq.~(\ref{4.1eq4}), we use the symmetrical negative cosine similarity as the metric of distance.

Finally, the training objective $\mathcal{L}$ is formulated as:
\begin{equation}
    \mathcal{L} = \underbrace{(\lambda_1\mathcal{L}_{unmix}+\lambda_2\mathcal{L}_{mix})}_{sup}+
    \underbrace{(\lambda_3\mathcal{L}_{con\text{-}g}+ \lambda_4\mathcal{L}_{con\text{-}l})}_{unsup},
\label{eq:final}
\end{equation}
where $\lambda_{1},\lambda_2,\lambda_3,\lambda_4$ are hyperparameters to leverage the relative importance of different loss components. 


\section{Experiments}

\subsection{Data and evaluation metric}
\begin{figure}[t]
\centering
\includegraphics[width=0.9\linewidth]{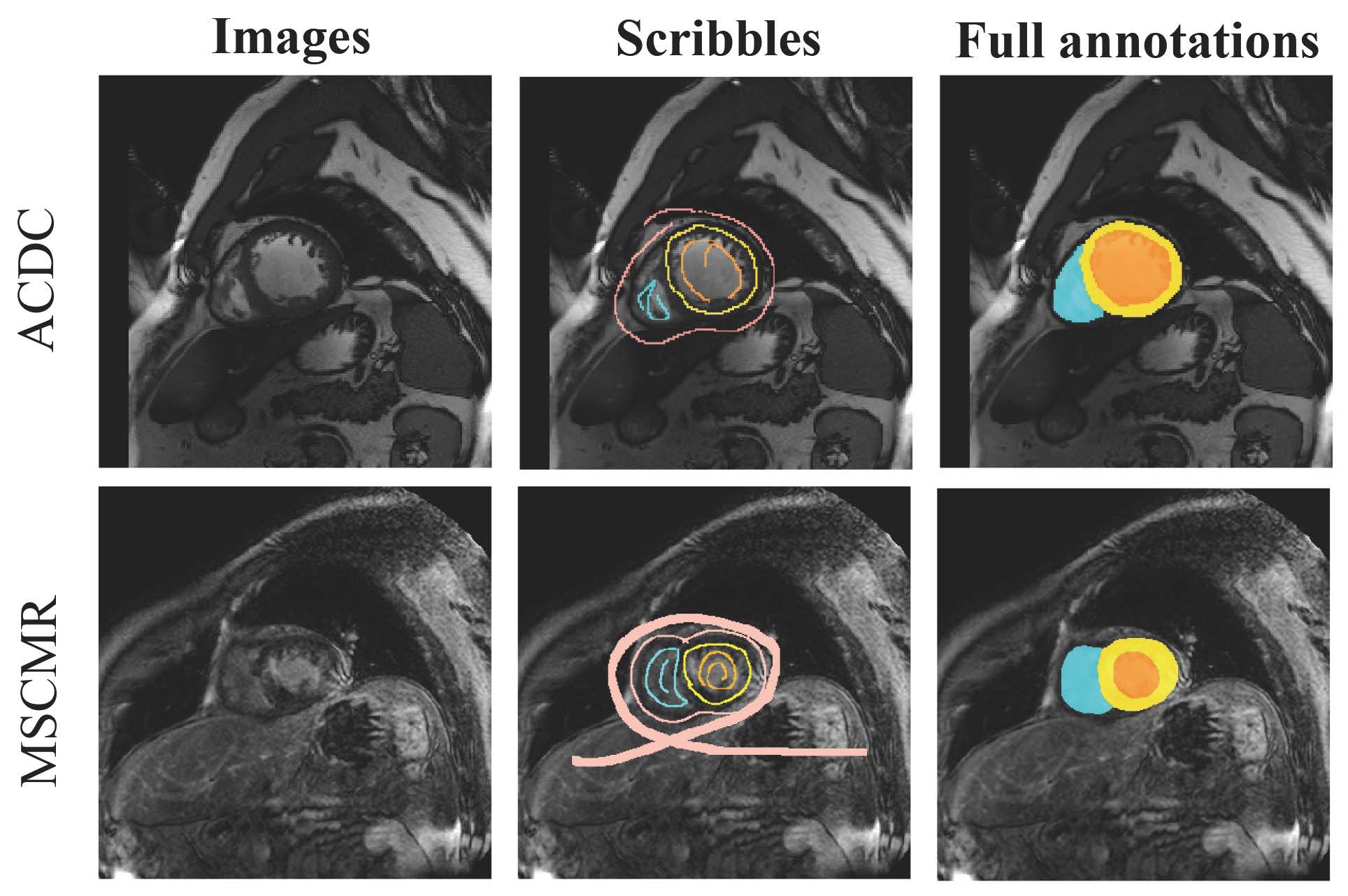}
\caption{Examples from ACDC and MSCMRseg datasets. Since boundaries of structures in MSCMRseg images are general more difficult to distinguish, a thicker scribble is included to annotate the background for more supervision.}
\label{fig:scribble}
\end{figure}

CycleMix is evaluated on two open datasets, \textit{i.e.}, ACDC and MSCMRseg, on which rich results have been reported in literature for comparisons. In addition, we use ACDC dataset for extensive parameter studies.

 
\textbf{ACDC}\cite{8360453} dataset is composed of 2-dimensional cine-MRI images from 100 patients. 
The cine-MRI images were obtained using two MRI scanners of various magnetic strengths and different resolutions.
For each patient, manual annotations of right ventricle (RV), left ventricle (LV) and myocardium (MYO) are provided for both the end-diastolic (ED) and end-systolic (ES) phase.
Following \cite{9389796}, the 100 subjects in ACDC dataset is randomly divided into 3 sets of 70 (training), 15( validation), 15 (test) subjects for experiments.
To compare with the previous state-of-the-art methods, which use unpaired masks to learn shape priors,
we further divided the training set into two halves,  35 training images with scribble labels and 35 mask images with heart segmentation.
\emph{Unless specified, we only used 35 training images when training the proposed CycleMix and baselines}.

\textbf{MSCMRseg}\cite{8458220,Zhuang2016MultivariateMM} contains late gadolinium enhancement (LGE) MRI images collected from 45 patients who underwent cardiomyopathy, which represents more challenges for automatic segmentation than the unenhanced cardiac MRI.
Gold standard segmentation of LV, MYO, RV of these images has also been released by the organizers.
Following \cite{yue2019cardiac}, we randomly divided the images from 45 patients into 3 sets, including 25 for training, 5 for validation and 20 for test. 

\textbf{Scribble annotations}. For ACDC dataset, we used the released expert-made scribble annotations \cite{9389796}. 
To obtain realistic scribble annotations, we further manually annotate the MSCMRseg dataset, following the principles in \cite{9389796}.
The average image coverages of scribbles for background, RV, MYO, LV are 3.4\%, 27.7\%, 31.3\%, and 24.1\%, respectively.
Figure.~\ref{fig:scribble} presents two exemplar images and their annotations from the two datasets.
Please refer to supplementary material for more details of scribble annotations.

\textbf{Evaluation}. We adopted the Dice coefficient~\cite{10.2307/1932409} to evaluate the performance of each method, which gauges the similarity of two segmentation masks.
\begin{figure*}[t]
\centering
\includegraphics[width=0.9\linewidth]{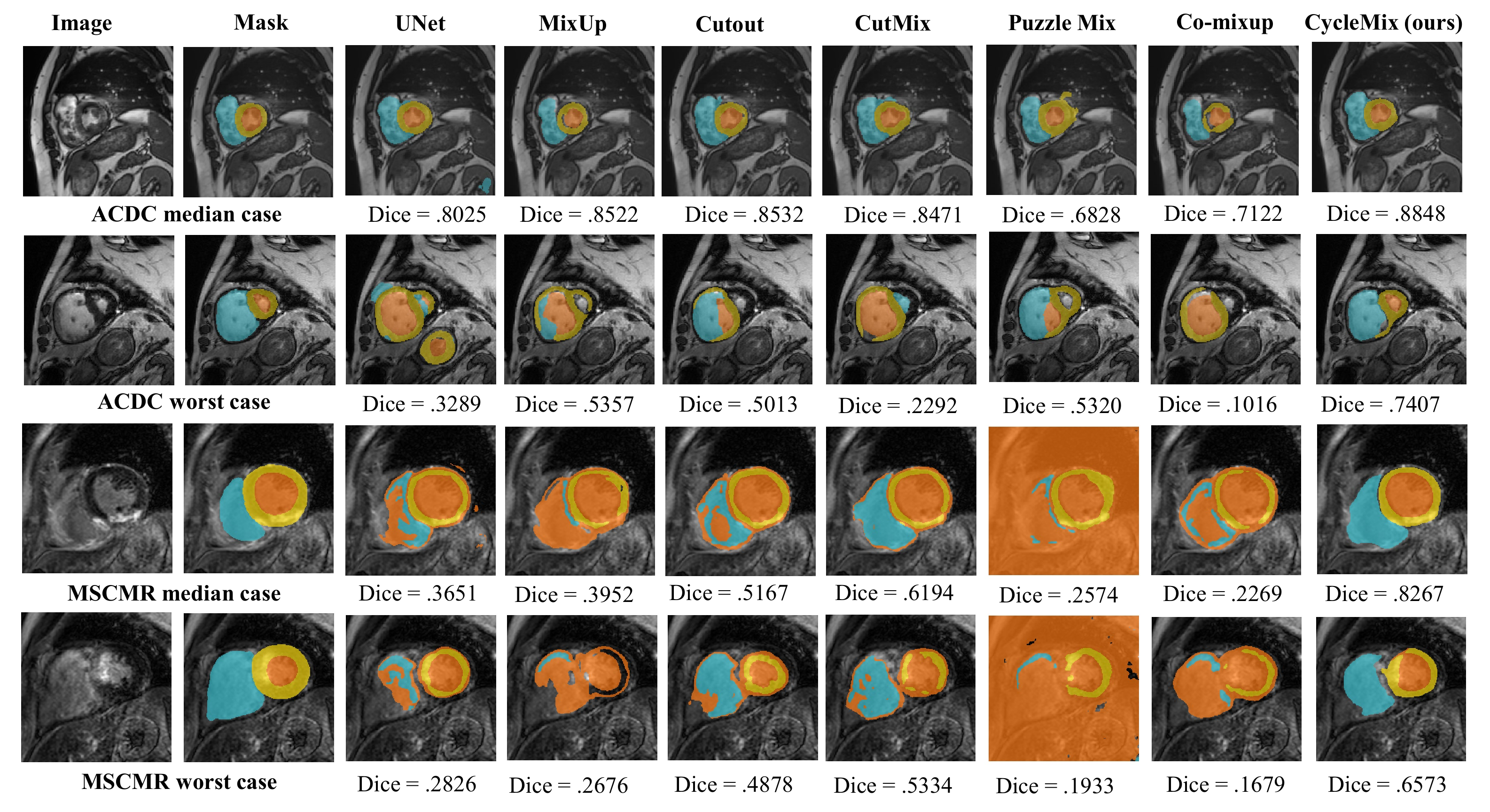}
\caption{Quantitative comparison of the proposed method on ACDC and MSCMRseg datasets. The selected subjects were the median and the worst cases with regard to the Dice scores of the results of fully-supervised segmentation by UNet$^+_\text{F}$. 
} 
\label{fig:visual}
\end{figure*}
\subsection{Experimental setup}
\textbf{Implementation Details}. We adopted the 2D variant of UNet~\cite{baumgartner2017exploration}, denoted as UNet$^+$, as the network architecture of CylceMix for all experiments, which was implemented using Pytorch. 
Since the provided images have different resolutions, We first resampled them and their annotations into a common in-plane resolution of $1.37\times1.37$ mm.
Then, all images were cropped or padded to the same image size of $212\times212$ pixel.
During training, we normalized the intensity of each image to zero mean and unit variance.
The learning rate was fixed to 0.0001.
We empirically set $\lambda_1=\lambda_2=\lambda_4=1$ and $\lambda_3=0.05$ in Eq.(\ref{eq:final}).
All models were trained using one single NVIDIA 3090Ti 24GB GPU for 1000 epochs.

\textbf{Baseline settings}. The proposed CycleMix was trained with \textit{scribble annotations}. 
Firstly, we compared it with baselines trained on scribble-annotated datasets.
Recently, there are several works leveraged GAN networks to learn shape priors. 
We also compared with these challenging benchmarks which require \textit{extra unpaired segmentation masks} to train GAN networks.
Finally, we considered several \textit{supervised methods} as upper bounds, which were trained on fully-annotated datasets.
\begin{itemize}
    \item \textit{Baselines}: 
    We first compared to UNet$^+_{pce}$ trained with cross entropy loss of annotated pixels in \cite{Tang2018OnRL}.
    Then, we applied different mix-up augmentation strategy to UNet$^+_{pce}$, \textit{i.e.}, MixUp~\cite{zhang2018mixup}, CutMix~\cite{yun2019cutmix}, Puzzle Mix~\cite{kimICML20}, Co-mixup~\cite{kim2021comixup}.
    Finally, we included the experiment results on ACDC dataset reported in \cite{9389796} for reference, \textit{i.e.,} UNet$_{pce}$\cite{tang2018normalized},UNet$_{wpce}$\cite{9389796}, UNet$_{CRF}$~\cite{zheng2015conditional}.

    \item \textit{Challenging benchmarks}: 
    The above baselines do not leverage additional unpaired segmentation masks during training. 
    For more challenging benchmarks, we compared with four works using extra unpaired data to learn shape priors, including PostDAE~\cite{Larrazabal2020PostDAEAP}, UNet$_D$~\cite{9389796}, ACCL~\cite{zhang2020accl}, MAAG~\cite{9389796}.
    We refered to their segmentation results reported in ~\cite{9389796} on ACDC dataset for comparison.
    
    \item \textit{Supervised methods}: Finally, we performed the comparison in fully-supervised segmentation. 
    Firstly, we applied UNet$^+$ in ~\cite{baumgartner2017exploration} to the training data of full annotations using conventional cross entropy loss, referred to as UNet$^+_F$.
    Then, we applied Puzzle Mix augmentation strategy to UNet$^+_F$, and obtained the Puzzle Mix$_F$.
    Finally, we trained CycleMix with fully annotated data, denoted as CycleMix$_F$, and compared with UNet$^+_F$ and Puzzle Mix$_F$ on both ACDC and MSCMRseg datasets. 
\end{itemize}

\subsection{Comparison with different mix-up strategies}
\begin{table*}[htb]
	\caption{The performance (Dice Scores) on ACDC and MSCMRseg dataset of CycleMix compared with different mixup strategies. \textbf{Bold} denotes the best performance, \underline{underline} denotes the second best performance.}\label{tab:tab1}
	\centering
		\resizebox{0.9\linewidth}{!}{
			\begin{tabular}{cccccccccc}
				\toprule
				\multirow{2}{*}{Methods}&\multirow{2}{*}{Data}&\multicolumn{4}{c}{ACDC}& \multicolumn{4}{c}{MSCMRseg}\\
				\cmidrule(lr){3-6}\cmidrule(lr){7-10}
				&&\multicolumn{1}{c}{LV} & MYO & RV & \multicolumn{1}{c|}{Avg} &LV & MYO & RV&\multicolumn{1}{c}{Avg}\\
				\midrule
				\multicolumn{6}{l}{35 scribbles}\\
				\midrule
				\multicolumn{1}{l|}{UNet$^+_{pce}$}&\multicolumn{1}{c|}{scribbles}&.785$\pm$.196&.725$\pm$.151&.746$\pm$.203&\multicolumn{1}{c|}{.752}&.494$\pm$.082&.583$\pm$.067&.057$\pm$.022&.378\\ 
				\multicolumn{1}{l|}{MixUp\cite{zhang2018mixup}}&\multicolumn{1}{c|}{scribbles}&.803$\pm$.178&.753$\pm$.116&.767$\pm$.226&\multicolumn{1}{c|}{.774}&.610$\pm$.144&.463$\pm$.147&.378$\pm$.153&.484\\
				\multicolumn{1}{l|}{Cutout\cite{devries2017cutout}}&\multicolumn{1}{c|}{scribbles}&.832$\pm$.172&.754$\pm$.138&.812$\pm$.129&\multicolumn{1}{c|}{.800}&.459$\pm$.077&.641$\pm$.136&.697$\pm$.149&.599\\
				\multicolumn{1}{l|}{CutMix\cite{yun2019cutmix}}&\multicolumn{1}{c|}{scribbles}&.641$\pm$.359&.734$\pm$.144&.740$\pm$.216&\multicolumn{1}{c|}{.705}&.578$\pm$.063&.622$\pm$.121&\underline{.761$\pm$.105}&.654\\
				\multicolumn{1}{l|}{Puzzle Mix\cite{kimICML20}}&\multicolumn{1}{c|}{scribbles}&.663$\pm$.333&.650$\pm$.231&.559$\pm$.343&\multicolumn{1}{c|}{.624}&.061$\pm$.021&.634$\pm$.084&.028$\pm$.012&.241\\
				\multicolumn{1}{l|}{Co-mixup\cite{kim2021comixup}}&\multicolumn{1}{c|}{scribbles}&.622$\pm$.304&.621$\pm$.214&.702$\pm$.211&\multicolumn{1}{c|}{.648}&.356$\pm$.075&.343$\pm$.067&.053$\pm$.022&.251\\
			    \multicolumn{1}{l|}{CycleMix(ours)}&\multicolumn{1}{|c|}{scribbles}&\textbf{.883$\pm$.095}&\underline{.798$\pm$.075}&\underline{.863$\pm$.073}&\multicolumn{1}{c|}{\textbf{.848}}&\textbf{.870$\pm$.061}&\underline{.739$\pm$.049}&\textbf{.791$\pm$.072}&\textbf{.800}\\
				\midrule
				\multicolumn{6}{l}{35 masks}\\
				\midrule
			    \multicolumn{1}{l|}{UNet$^+_F$}&\multicolumn{1}{|c|}{masks}&.849$\pm$.152&.792$\pm$.140&.817$\pm$.151&\multicolumn{1}{c|}{.820}&.857$\pm$.055&.720$\pm$.075&.689$\pm$.120&.755\\
			    \multicolumn{1}{l|}{Puzzle Mix$_F$\cite{kimICML20}}&\multicolumn{1}{|c|}{masks}&\underline{.849$\pm$.182}&\textbf{.807$\pm$.088}&\textbf{.865$\pm$.089}&\multicolumn{1}{c|}{\underline{.840}}&\underline{.867$\pm$.042}&\textbf{.742$\pm$.043}&.759$\pm$.039&\underline{.789}\\
				\toprule
		\end{tabular}}
\end{table*}
Table.~\ref{tab:tab1} presents the performance of CycleMix on ACDC and MSCMRseg datasets.
We compared with different data augmentation methods, \textit{i.e.}, Mixup, Cutout, CutMix, Puzzle Mix, Comix-up as strong baselines. 
Here, we used 35 subjects for training, and the results using 70 training images are presented in supplementary material. 

When only scribble annotations are available, Puzzle Mix achieved poor performance, with average Dice Scores of $62.4\%$ on ACDC dataset and $24.1\%$ on MSCMRseg. When with our proposed augmentation and regularization of supervision, CycleMix boosted the performance to reach Dice of $84.8\%$ and $80.0\%$ for the two datasets, respectively,
demonstrating improvements of $22.4\%$ and $55.9\%$.

Furthermore, the average Dice Score of CycleMix not only surpassed all weakly-supervised baselines by a large margin, but also exceeded the two fully-supervised methods.
Particularly on the challenging task of MSCMRseg dataset, CycleMix achieved average Dice 0.800, with $14.6\%$ increment than CutMix which ranks the second in the scribble supervision leader board.
For the fully-supervised methods, 
one can observe CycleMix (marginally) outperformed both UNet$_F$ and Puzzle Mix$_F$ in Table.~\ref{tab:tab1}.
Specifically, CycleMix with scribble supervision obtained an average improvement of $0.8\%$ (84.8\% vs 84.0\%) and $1.1\%$ (80.0\% vs 78.9\%) on MSCMRseg and ACDC dataset, respectively.

Figure.~\ref{fig:visual} visualizes results on the worst and median cases selected using the fully-supervised UNet.
It is observed that Puzzle Mix could fail in the scribble supervision-based segmentation, especially on the challenging task of MSCMRseg.
This may be due to its transportation strategy of image patches, which is more likely to change the shape of the target structure than other mix-up strategies based on linear interpolation or local replacement. Similar behavior could be seen from Co-mixup which adopts the similar transportation strategy to that of Puzzle Mix.
Therefore, it is more difficult for the segmentation model to learn the shape priors, especially in the case of a small training dataset.
CycleMix overcomes this disadvantage by combining losses of mixed images and unmixed images, \textit{i.e.}, $\mathcal{L}_{mix}$ and  $\mathcal{L}_{unmix}$, and leveraging consistency regularization  to preserve shape priors, 
which will be further explored in the ablation study.

\subsection{Comparison with weakly-supervised methods}
Table.~\ref{tab2} presents the results on the ACDC dataset. 
The previous best method, MAMG~\cite{9389796} exploited the unpaired masks from 35 additional subjects, and achieved 81.6\% Dice Score with the assistance of multi-scale GAN. 
Without  these masks,  CycleMix still achieved a new state-of-the-art (SOTA)  Dice of 84.8\% average, with a promising margin over MAMG. 
For the RV structure with more shape variation, CycleMix obtained remarkable gains of 11.1\% over MAMG (86.3\% vs 75.2\%).
For the other methods, CycleMix demonstrated more significant performance improvements.
We concluded that despite the additional masks, the models could learn very limited prior shapes through GAN when the number of training images is small.
Thanks to the mix augmentation and consistency regularization for scribble supervision, CycleMix learned robust shape priors and set a new SOTA of segmentation.

Moreover, as one can observed from the upper part of Table.~\ref{tab2},  CycleMix consistently outperformed all the other scribble supervision-based methods.
Particularly, CycleMix obtained average performance gain up to $8.2\%$ than UNet$_{pce}$ which ranks the 2nd. 

\begin{table}[t]
	\caption{The performance (Dice Scores) on ACDC dataset of CycleMix compared with state-of-the-art weakly-supervised methods. We referred to their segmentation results reported in ~\cite{9389796} on ACDC dataset for comparison.}\label{tab2}
	\centering
		\resizebox{1\linewidth}{!}{
			\begin{tabular}{l|ccccc}
				\toprule
				Methods& Data& LV & MYO & RV &\multicolumn{1}{|c}{Avg}\\
				\midrule
				\multicolumn{6}{l}{35 scribbles}\\
				\midrule
				\multicolumn{1}{l|}{UNet$_{pce}$\cite{tang2018normalized}}&\multicolumn{1}{c|}{scribbles}&.842&.764&.693&\multicolumn{1}{|c}{.766}\\
				\multicolumn{1}{l|}{UNet$_{wpce}$\cite{9389796}}&\multicolumn{1}{c|}{scribbles}&.784&.675&.563&\multicolumn{1}{|c}{.674}\\
				\multicolumn{1}{l|}{UNet$_{CRF}$\cite{zheng2015conditional}}&\multicolumn{1}{|c|}{scribbles}&.766&.661&.590&\multicolumn{1}{|c}{.672}\\
			    \multicolumn{1}{l|}{CycleMix(ours)}&\multicolumn{1}{|c|}{scribbles}&\textbf{.883}&\underline{.798}&\textbf{.863}&\multicolumn{1}{|c}{\textbf{.848}}\\
				\midrule
				\multicolumn{6}{l}{35 scribbles + 35 unpaired masks}\\
				\midrule
				\multicolumn{1}{l|}{UNet$_{D}$\cite{9389796}}&\multicolumn{1}{|c|}{scribbles+masks}&.404&.597&\underline{.753}&\multicolumn{1}{|c}{.585}\\ \multicolumn{1}{l|}{PostDAE\cite{Larrazabal2020PostDAEAP}}&\multicolumn{1}{|c|}{scribbles+masks}&.806&.667&.556&\multicolumn{1}{|c}{.676}\\   \multicolumn{1}{l|}{ACCL\cite{zhang2020accl}}&\multicolumn{1}{|c|}{scribbles+masks}&.878&.797&.735&\multicolumn{1}{|c}{.803}\\
			    \multicolumn{1}{l|}{MAAG\cite{9389796}}&\multicolumn{1}{|c|}{scribbles+masks}&\underline{.879}&\textbf{.817}&.752&\multicolumn{1}{|c}{\underline{.816}}\\
				\toprule
		\end{tabular}}
\end{table}

\subsection{Ablation study} 

	\begin{table*}[!t]
		\caption{Ablation study: CycleMix for image segmentation with different settings, including loss of unmixed samples ($\mathcal{L}_{unmix}$), loss of mixed samples ($\mathcal{L}_{mix}$), global consistency loss ($\mathcal{L}_{con\text{-}g}$), random occlusion ($\mathbbm{1}_O$), local consistency loss ($\mathcal{L}_{con\text{-}l}$).
		Symbol * indicates statistically significant improvement given by a Wilcoxon signed-rank test with $p\leq0.05$.
		}\label{tab3}
		\begin{center}
			\resizebox{0.85\linewidth}{!}{
				\begin{tabular}{c|ccccc|ccc|c}
					\toprule
					Methods & $\mathcal{L}_{unmix}$ &$\mathcal{L}_{mix}$& $\mathcal{L}_{con\text{-}g}$ & $\mathbbm{1}_O$ & $\mathcal{L}_{con\text{-}l}$ & LV & MYO & RV & Avg \\
					\midrule
					\multicolumn{1}{c|}{\#1}&$\checkmark$&$\times$&$\times$&$\times$&\multicolumn{1}{c|}{$\times$}&.785$\pm$.196&.725$\pm$.151&.746$\pm$.203&\multicolumn{1}{|c}{.752}\\
					\multicolumn{1}{c|}{\#2}&$\checkmark$&$\checkmark$&$\times$&$\times$&\multicolumn{1}{c|}{$\times$}&.863$\pm$.104$^*$&.783$\pm$.086$^*$&.782$\pm$.173&\multicolumn{1}{|c}{.809$^*$}\\

					\multicolumn{1}{c|}{\#3}&$\checkmark$&$\checkmark$&$\checkmark$&$\times$&\multicolumn{1}{c|}{$\times$}&.867$\pm$.130&.786$\pm$.114&\multicolumn{1}{c|}{.837$\pm$.097$^*$}&\multicolumn{1}{|c}{.830$^*$}\\
					\multicolumn{1}{c|}{\#4}&$\checkmark$&$\checkmark$&$\checkmark$&$\checkmark$&\multicolumn{1}{c|}{$\times$}&\textbf{.898$\pm$.059$^*$}&.786$\pm$.078&\multicolumn{1}{c|}{.847$\pm$.132$^*$}&\multicolumn{1}{|c}{.843$^*$}\\
					\multicolumn{1}{c|}{\#5}&$\checkmark$&$\checkmark$&$\checkmark$&$\checkmark$&\multicolumn{1}{c|}{$\checkmark$}&.883$\pm$.095&\textbf{.798$\pm$.075$^*$}&\textbf{.863$\pm$.073}&\multicolumn{1}{|c}{\textbf{.848}}\\
					\toprule
			\end{tabular}}
		\end{center}
	\end{table*}  

This section studies the effectiveness of our proposed strategies,
including the usage of unmix loss ($\mathcal{L}_{unmix}$), mixed loss ($\mathcal{L}_{mix}$), global consistency loss ($\mathcal{L}_{con\text{-}g}$), random occlusion ($\mathbbm{1}_O$), and local consistency loss ($\mathcal{L}_{con\text{-}l}$).
Table.~\ref{tab3} presents the details.

\noindent
\textbf{Effectiveness of global consistency}: UNet$^+$ (\#1) with cross entropy loss of annotated pixels could achieve the average Dice Score of 75.2\%. 
When we added mixed loss $\mathcal{L}_{mix}$ as additional segmentation loss, the average performance increased by $5.7\%$  (75.2\% to 80.9\%);
and when the global consistency ($\mathcal{L}_{con\text{-}g}$) was included for regularization, the average Dice was further boosted to 83.0\%. 
This was attribute to the fact that the combination of global consistency could encourage segmentation model to learn the mix-invariant property, and enhance the ability of model to learn robust shape priors.

\noindent
\textbf{Effectiveness of random occlusion}: For model \#4, we observed that random occlusion ($\mathbbm{1}_O$) brought a convincing average Dice Score improvement of $1.3\%$ ($84.3\%$ vs $83.0\%$), demonstrating its effectiveness to enhance the localization ability of model via additional augmentation of scribble supervision.

\noindent
\textbf{Effectiveness of local consistency}: When local consistency  ($\mathcal{L}_{con\text{-}l}$) was adopted for shape regularization, model \#5 performed marginally better than model \#4, with an increase of $0.8\%$ average Dice Score (84.8\% vs 84.0\%).
Particularly on MYO structure, $ \mathcal{L}_{con\text{-}l}$ helped obtaining a statistically significant improvement of $1.2\%$ Dice, indicating the benefit of local consistency in shape regularization for segmentation of challenging structures. 

\subsection{Data sensitivity study}
\begin{table}[t]
	\caption{Data sensitivity study: the performance of CycleMix with different ratio of scribbles to full annotations.}\label{tab4}
	\centering
		\resizebox{1\linewidth}{!}{
			\begin{tabular}{c|c|ccc|c}
				\toprule
				Methods& scribble: full& LV & MYO & RV &\multicolumn{1}{|c}{Avg}\\
				\midrule
				\multicolumn{1}{c|}{1}&\multicolumn{1}{c|}{35:00}&.883$\pm$.095&.798$\pm$.075&.863$\pm$.073&\multicolumn{1}{|c}{.848}\\
				\midrule
				\multicolumn{1}{c|}{2}&\multicolumn{1}{c|}{70:00}&.880$\pm$.115&.825$\pm$.072&.860$\pm$.089&\multicolumn{1}{|c}{.855}\\
				\multicolumn{1}{c|}{3}&\multicolumn{1}{c|}{56:14}&.898$\pm$.075&.842$\pm$.072&.876$\pm$.112&\multicolumn{1}{|c}{.872}\\
				\multicolumn{1}{c|}{4}&\multicolumn{1}{c|}{42:28}&.911$\pm$.063&.854$\pm$.056&.883$\pm$.076&\multicolumn{1}{|c}{.883}\\
				\multicolumn{1}{c|}{5}&\multicolumn{1}{c|}{28:42}&.902$\pm$.080&.851$\pm$.065&\textbf{.899$\pm$.058}&\multicolumn{1}{|c}{.884}\\
			    \multicolumn{1}{c|}{6}&\multicolumn{1}{|c|}{14:56}&.906$\pm$.065&.856$\pm$.066&.893$\pm$.083&\multicolumn{1}{|c}{.885}\\
			    \multicolumn{1}{c|}{7}&\multicolumn{1}{|c|}{00:70}&\textbf{.919$\pm$.065}&.\textbf{858$\pm$.058}&.882$\pm$.088&\multicolumn{1}{|c}{\textbf{.886}}\\
				\midrule
			    \multicolumn{1}{c|}{UNet$^+_F$}&\multicolumn{1}{|c|}{00:70}&.883$\pm$.130&.831$\pm$.093&.870$\pm$.096&\multicolumn{1}{|c}{.862}\\
				\toprule
		\end{tabular}}
\end{table}	

This study investigates the performance of CycleMix with different training images of scribble annotation and full annotation. For this study, we included all the 70 training images from ACDC and altered the ratio between the two sets of annotations. Table.~\ref{tab4} presents the results.


Interestingly, one can observe that when the ratio of full annotation reaches 20\% (56:14),
CycleMix outperformed the fully-supervised UNet$_F^+$ by a margin of $1.0\%$ (87.2\% vs 86.2\%) on the average Dice.          
As expected, the performance of CycleMix tended to increase as the ratio of fully-annotated subjects increases.
One can observe that the general performance of CycleMix converged when the ratio of fully-annotated data reaches 40\%. 
This confirms that CycleMix could achieve a satisfactory segmentation result with a relatively small amount of full annotations.

\subsection{Experiments on fully-annotated data}
\begin{table}[t]
	\caption{Comparisons on fully-supervised segmentation.}\label{tab6}
\centering
\resizebox{\linewidth}{!}{
\begin{tabular}{l|ccc|c}
	\toprule
	Methods& LV & MYO & RV &\multicolumn{1}{c}{Avg}\\
	\midrule
	\multicolumn{5}{l}{ACDC dataset} \\
	\midrule
\multicolumn{1}{l|}{UNet$^+_F$}&.883$\pm$.130&.831$\pm$.093&.870$\pm$.096&\multicolumn{1}{|c}{.862}\\
\multicolumn{1}{l|}{Puzzle Mix$_F$\cite{kimICML20}}&.912$\pm$.082&.842$\pm$.081&\textbf{.887$\pm$.066}&\multicolumn{1}{|c}{.880}\\
\multicolumn{1}{l|}{CycleMix$_F$}&\textbf{.919$\pm$.065}&\textbf{.858$\pm$.058}&.882$\pm$.088&\multicolumn{1}{|c}{\textbf{.886}}\\
	\midrule
	\multicolumn{5}{l}{MSCMRseg dataset}\\
	\midrule
\multicolumn{1}{l|}{UNet$^+_F$}&.857$\pm$.055&\.720$\pm$.075&.689$\pm$.120&\multicolumn{1}{|c}{.755}\\
\multicolumn{1}{l|}{Puzzle Mix$_F$\cite{kimICML20}}&\textbf{.867$\pm$.042}&.742$\pm$.043&.759$\pm$.039&\multicolumn{1}{|c}{.789}\\
\multicolumn{1}{l|}{CycleMix$_F$}&.864$\pm$.034&\textbf{.785$\pm$.042}&\textbf{.781$\pm$.066}&\multicolumn{1}{|c}{\textbf{.810}}\\
	\toprule
\end{tabular}
}

\end{table}
Table.~\ref{tab6} provides the Dice Score of fully-supervised segmentation on ACDC and MSCMRseg datasets. 
With fully-annotated labels, Puzzle Mix demonstrated competitive performance, improving the average Dice  of Unet$_F^+$ from $86.2\%$ to $88.0\%$ on ACDC, and from $75.5\%$ to $78.9\%$ on MSCMRseg.
By contrast, CycleMix could improve more,  but the margins were not so exciting as it did in the scribble supervision.
This indicates that  CycleMix can excel in both scribble supervision-based and fully-supervised segmentation, but its advantage could be more evident in the former applications, for which CycleMix has been specifically designed.  
\section{Conclusions}
In this paper, we have investigated a novel weakly-supervised learning framework, CycleMix, to learn segmentation from scribble supervision.
The proposed method utilizes mix augmentation of supervision and cycle consistency of segmentation to enhance the generalization ability of segmentation models.
CycleMix was evaluated on two open datasets, \textit{i.e.}, ACDC and MSCMRseg, and achieved the new state-of-the-art performance.

{
    \small
    \bibliographystyle{ieee_fullname}
    \bibliography{macros,main}
}




\end{document}